\documentclass[journal,10pt]{IEEEtran}

\usepackage{graphicx}
\graphicspath{{figures/}}
\usepackage{tikz}

\usepackage{amsmath,amsfonts,stmaryrd,dsfont}
\usepackage{multirow}

\usepackage{algorithm,algorithmic,setspace}

\usepackage{array}

\usepackage{url}
\urldef{\urldemos}\url{http://www.perso.telecom-paristech.fr/%7Emagron/ressources/levyNMF.m}

\begin{document}

\title{L\'evy NMF for robust nonnegative source separation}

\author{Paul~Magron,
	Roland~Badeau,~\IEEEmembership{Senior Member,~IEEE,}
	and~Antoine~Liutkus,~\IEEEmembership{Member,~IEEE}% <-this % stops a space
	\thanks{P. Magron and R. Badeau are with LTCI, CNRS, T\'{e}l\'{e}com ParisTech, Universit\'{e} Paris-Saclay, 75013, Paris, France. A. Liutkus is with Inria, Nancy Grand-Est, Multispeech team, LORIA UMR 7503, France (e-mail: firstname.lastname@\{telecom-paristech,inria\}.fr). \newline This work was partly supported by the research programme KAMoulox (ANR-15-CE38-0003-
01) funded by ANR, the French State agency for research.}
}

\markboth{IEEE SIGNAL PROCESSING LETTERS}%
{Shell \MakeLowercase{\textit{et al.}}: PasNMF}

\maketitle

\begin{abstract}
Source separation, which consists in decomposing data into meaningful structured components, is an active research topic in many areas, such as music and image signal processing, applied physics and text mining. In this paper, we introduce the Positive $\alpha$-stable (P$\alpha$S) distributions to model the latent sources, which are a subclass of the stable distributions family. They notably permit us to model random variables that are both nonnegative and impulsive. Considering the L\'evy distribution, the only P$\alpha$S distribution whose density is tractable, we propose a mixture model called L\'evy Nonnegative Matrix Factorization (L\'evy NMF). This model accounts for low-rank structures in nonnegative data that possibly has high variability or is corrupted by very adverse noise. The model parameters are estimated in a maximum-likelihood sense. We also derive an estimator of the sources given the parameters, which extends the validity of the generalized Wiener filtering to the P$\alpha$S case. Experiments on synthetic data show that L\'evy NMF compares favorably with state-of-the art techniques in terms of robustness to impulsive noise. The analysis of two types of realistic signals is also considered: musical spectrograms and fluorescence spectra of chemical species. The results highlight the potential of the L\'evy NMF model for decomposing nonnegative data.
\end{abstract}

\begin{IEEEkeywords}
L\'evy distribution, Positive alpha-stable distribution, nonnegative matrix factorization, source separation.
\end{IEEEkeywords}

\IEEEpeerreviewmaketitle

\section{Introduction}

\IEEEPARstart{S}{ource} separation consists in extracting underlying components called \textit{sources} that add up to form an observable signal called \textit{mixture}. This issue occurs in various fields such as data mining~\cite{Pauca2004}, face recognition~\cite{Guillamet2002} or applied physics~\cite{Sajda2004}.
A groundbreaking idea presented in~\cite{Lee1999} is to exploit the fact that the observations are often nonnegative, so that they should be decomposed as a sum of only nonnegative terms. This Nonnegative Matrix Factorization (NMF) has shown successful in many fields such as audio signal processing~\cite{Smaragdis2003}, computer vision~\cite{Lee1999}, spectroscopy~\cite{Liu2013} and many others.

NMF, originally introduced as a rank-reduction method, approximates a nonnegative data matrix $X$ as a product of two low-rank nonnegative matrices $W$ and $H$. The factorization can be obtained by optimizing a cost function measuring the error between $X$ and $WH$, such as the Euclidean, Kullback-Leibler (KL~\cite{Lee1999}) and Itakura-Saito (IS~\cite{Fevotte2009}) cost functions. This may often be framed in a probabilistic framework, where the cost function appears as the negative log-likelihood of the data, e.g. Gaussian~\cite{Fevotte2009,Schmidt2008} or Poisson~\cite{Virtanen2008,Cemgil2009}. Addressing the estimation problem in a Maximum A Posteriori (MAP) sense makes it possible to incorporate some prior distribution over the parameters $W$ and $H$. This allows some regularization schemes enforcing desirable properties for the parameters such as harmonicity, temporal smoothness~\cite{Bertin2010} or sparsity~\cite{Virtanen2007}.

However, the above-mentioned distributions fail to provide good results when the data is very impulsive or contains outliers. This comes from their rapidly decaying tails, that cannot account for really unexpected observations.
The family of heavy-tailed \textit{stable} distributions~\cite{Samoradnitsky1994} was thus found useful for robust signal processing~\cite{Godsill1999,Bassiou2013}.
%The fact is that the family of heavy-tailed \textit{stable} distributions~\cite{Samoradnitsky1994} has been considered in many applications and was found useful for robust signal processing~\cite{Godsill1999,Bassiou2013}.
A subclass of this family, called the Symmetric $\alpha$-stable (S$\alpha$S) distributions, has been used in audio~\cite{Liutkus2015,Liutkus2015a} for modeling complex Short-Term Fourier Transforms (STFT).
%Based on these findings, an extension of the ISNMF was presented in~\cite{}, where the NMF now pertains to the dispersion parameters of Cauchy distributed sources.
An estimation framework based on the Markov Chain Monte Carlo (MCMC) method has been proposed~\cite{Simsekli2015} to perform the separation of S$\alpha$S mixtures. The common ground of these methods is to assume all observations as independent but to impose low-rank constraints on the nonnegative \textit{dispersion parameters} of the sources, and not on their actual outcomes.
%,that are taken as symmetrically distributed around $0$.

In this paper, we address the problem of modeling and separating nonnegative sources from their mixture, while still constraining their dispersion to follow an NMF model. To do so, we consider another subclass of the stable family which models nonnegative random variables: the positive $\alpha$-stable (P$\alpha$S) distributions. They also benefit from being heavy-tailed and are thus expected to yield robust estimates. Since the Probability Density Function (PDF) of those P$\alpha$S distributions does not admit a closed-form expression in general, we study more specifically the L\'evy case, which is a particular analytically tractable member of the family. We introduce the \emph{L\'evy NMF} model, where the dispersion parameters of the sources are structured through an NMF model and where realizations are necessarily nonnegative. The parameters are then estimated in a Maximum Likelihood (ML) sense by means of a Majorize-Minimization approach. We also derive an estimator of the sources which extends the validity of the generalized Wiener filtering to the P$\alpha$S case. Several experiments conducted on synthetic, audio and fluorescence spectroscopy signals show the potential of this model for a nonnegative source separation task and highlight its robustness to impulsive noise.

This paper is structured as follows. Section~\ref{sec:model} introduces the L\'evy NMF mixture model. Section~\ref{sec:estim} details the parameters estimation and presents an estimator of the sources. Section~\ref{sec:exp} experimentally demonstrates the denoising ability of the model and its potential in terms of source separation for both musical and chemometric applications.

\section{L\'evy NMF model}
\label{sec:model}

\subsection{Positive $\alpha$-stable distributions}

Stable distributions, denoted $\mathcal{S}(\alpha,\mu,\sigma,\beta)$, are heavy-tailed distributions parametrized by four parameters: a shape parameter $\alpha \in ]0;2]$ which determines the tails thickness of the distribution (the smaller $\alpha$, the heavier the tail of the PDF), a location parameter $\mu \in \mathbb{R}$, a scale parameter $\sigma \in ]0 ; + \infty [$ measuring the dispersion of the distribution around its mode,  and a skewness parameter $\beta \in [-1;1]$. The symmetric $\alpha$-stable (S$\alpha${S}) distributions, which are such that $\beta=0$, are an important subclass of the stable family and a growing topic of interest, notably in audio~\cite{Liutkus2015,Simsekli2015}.

Such distributions are said "stable" because of their additive property: a sum of $K$ independent stable random variables $X_k \sim \mathcal{S}(\alpha,\mu_k,\sigma_k,\beta)$ is also a stable random variable: $X = \sum_k X_k \sim \mathcal{S} \left( \alpha,\mu,\sigma ,\beta \right)$, with $\mu = \sum_k \mu_k$ and $\sigma^\alpha = \sum_k \sigma_k^\alpha$.

%However, stable distributions benefits from two important properties. First, they are heavy-tailed distribution (the lower $\alpha$, the heavier the tail), as illustrated in Fig.~\ref{fig:levy_gauss_poisson_log}.

% \begin{figure}
% 	\centering
% 	\includegraphics[scale=0.5]{levy_gauss_poisson_log}
% 	\caption{PDF of several distributions of scale parameter $1$: the L\'{e}vy distribution has an heavier tail than other popular distributions.}
% 	\label{fig:levy_gauss_poisson_log}
% \end{figure}

Stable distributions do not in general have a nonnegative support. However, it can be shown~\cite{Nolan2015} that when $\beta=1$ and $\alpha<1$, the support of the distribution is $[\mu;+\infty [$. In this paper, we consider that $\mu=0$ thus the support is $\mathbb{R}_+$. The Positive $\alpha$-stable distributions are therefore such that $\mathcal{P \alpha S}(\sigma) = \mathcal{S}(\alpha,0,\sigma,1)$ with $\alpha<1$.

\subsection{L\'evy NMF mixture model}

The only $\alpha$ for which the PDF of a P$\alpha$S distribution can be expressed in closed form is the L\'evy case $\mathcal{L}\left(\sigma\right)=P \frac{1}{2} S\left(\sigma\right)$:
%\footnote{Note that in general, the PDF of the L\'{e}vy distribution depends on a location parameter $\mu$. However, in this paper, we only consider P$\alpha$S distributions so we assume it is null.}:
\begin{equation}
p(x\mid\sigma) = 
\left\{
\begin{aligned}
& \sqrt{\frac{\sigma}{2 \pi}} \frac{1}{x^{3/2}} e^{-\frac{\sigma}{2x}}  & \text{if } x>0\\ 
& 0 & \text{ otherwise.}   \\
\end{aligned}
\right.
\end{equation}

We model all the entries $X\left(f,t\right)$ of a data matrix $X \in \mathbb{R}_+^{F \times T}$ as independent, and assume they are the sum of $K$ independent L\'{e}vy-distributed components $X_k(f,t) \sim \mathcal{L}(\sigma_k(f,t))$. Note that all entries are independent, which allows us to extend the standard notation to matrices. Then, $X \sim \mathcal{L}(\sigma)$ with $\sigma^{\odot 1/2} = \sum_k \sigma_k^{\odot 1/2}$, where $^{\odot}$ denotes the element-wise power.

The scale parameters are structured by means of an NMF~\cite{Lee1999}, which preserves the additive property of the model as in~\cite{Liutkus2015a}: $\sigma^{\odot 1/2} = WH$, with $W \in \mathbb{R}_+^{F \times K}$ and $H \in \mathbb{R}_+^{K \times T}$. We then refer to this model as the \emph{L\'evy NMF} model.
%since the scale parameters $\sigma(f,t)$ now pertain to L\'evy instead of Gaussian~\cite{Fevotte2009}, Cauchy~\cite{Liutkus2015a} or Poisson~\cite{Virtanen2008} random variables $X(f,t)$.

\section{Parameters estimation}
\label{sec:estim}

The parameters $\theta = \{ W,H \}$ are estimated in a Maximum Likelihood (ML) sense, which is natural in a probabilistic framework. The log-likelihood of the data is given by:
\begin{align*}
L(W,H) &= \sum_{f,t} \log(p(X(f,t);\sigma(f,t))) \\
& \overset{c}{=} \frac{1}{2} \sum_{f,t} \log([WH](f,t)^2) - \frac{[WH](f,t)^2}{X(f,t)} \\
& \overset{c}{=} -\frac{1}{2} d_{IS}([WH]^{\odot 2},X),
\end{align*}
where $\overset{c}{=}$ denotes equality up to an additive constant which does not depend on the parameters and $d_{IS}$ denotes the IS divergence~\cite{Fevotte2009}. We thus remark that maximizing the log-likelihood of the data in the L\'{e}vy NMF model is equivalent to minimizing the IS divergence between $[WH]^{\odot 2}$ and $X$, which boils down to minimizing the following cost function:
\begin{equation}
\mathcal{C}(W,H) = \sum_{f,t} \frac{[WH](f,t)^2}{X(f,t)} - 2 \log([WH](f,t)).
\label{eq:cost_levy}
\end{equation}

\subsection{Naive multiplicative updates}

The cost function~\eqref{eq:cost_levy} can be minimized with the same heuristic approach that has been pioneered in~\cite{Lee1999} and used in many NMF-related papers in the literature.
%We consider the cost function $\mathcal{C}(W,H) = \sum_{f,t} \frac{[WH](f,t)^2}{X(f,t)} - 2 \log([WH](f,t))$, which is equal to the IS divergence up to an additive constant.
The gradient of $\mathcal{C}$ with respect to a parameter $\theta$ ($W$ or $H$) is expressed as the difference between two nonnegative terms: $\nabla_{\theta} \mathcal{C}  = \nabla_{\theta}^+ - \nabla_{\theta}^-$, which leads to the multiplicative update rules (MUR):
\begin{equation}
{\theta} \leftarrow {\theta} \odot \frac{\nabla_{\theta}^-}{ \nabla_{\theta}^+} = {\theta} \odot a_{\theta},
\label{eq:mur_update}
\end{equation}
where $\odot$ (resp. the fraction bar) denotes the element-wise matrix multiplication (resp. division). For L\'evy NMF:
\begin{equation}
a_W = \frac{ [WH]^{\odot -1} H^T}{([WH] \odot X^{\odot -1}) H^T},
\label{eq:a_W}
\end{equation}
and
\begin{equation}
a_H = \frac{ W^T [WH]^{\odot -1} }{ W^T ([WH] \odot X^{\odot -1}) }.
\label{eq:a_H}
\end{equation}
Provided $W$ and $H$ have been initialized as nonnegative, they remain so throughout iterations. However, these updates do not guarantee a non-increasing cost function $\mathcal{C}$, which motivates the research for a novel optimization approach.
%These updates rules are derived straightforwardly with a classical nonnegative methodology, but they do not guarantee a non-increasing cost function $\mathcal{C}$. In practice, we observed that the cost function is not non-increasing, which motivates the research for a novel optimization approach.

% A possible way to overcome this limitation, and to increase the convergence rate, is to incorporate an exponent parameter $\eta_{\theta}$ in the update rule~\eqref{eq:mur_update}:

% \begin{equation}
% {\theta} \leftarrow {\theta} \times a_{\theta}^{\odot \eta_{\theta}}.
% \label{eq:mur_eta}
% \end{equation}
% %
% This strategy has been shown to provide good results in a classical NMF framework with $\beta$-divergences~\cite{Badeau2010}. However, in the L\'evy NMF framework, we do not have a method for choosing the optimal parameter $\eta_{\theta}$, which can vary over iterations.

\subsection{Majorize-Minimization updates}

An alternative way to derive update rules for estimating the parameters is to adopt a Majorize-Minimization (MM) approach~\cite{Hunter2004}. The core idea of this strategy is to find an auxiliary function $G$ which majorizes the cost function $\mathcal{C}$:
\begin{equation}
\forall (\theta,\overline{\theta}) \text{, } \mathcal{C}(\theta) \leq G(\theta,\overline{\theta}) \text{, and } \mathcal{C}(\overline{\theta}) = G(\overline{\theta},\overline{\theta}).
\label{eq:condMM}
\end{equation}
Then, given some current parameter $\overline{\theta}$, we aim at minimizing $G(\theta,\overline{\theta})$ in order to obtain a new parameter $\theta$. This approach guarantees that the cost function $\mathcal{C}$ will be non-increasing over iterations.
%Due to space constraints, we cannot provide the complete details of the MM updates derivation, but here are the key-points of the method:
Such an auxiliary function is obtained in a similar fashion as in~\cite{Fevotte2011,Fevotte2011a}. We have:
\begin{equation}
[WH](f,t)^2 =\left( \sum_k \rho_k(f,t) \frac{ W(f,k)H(k,t)}{\rho_k(f,t)} \right)^2,
\end{equation}
where
\begin{equation}
\rho_k(f,t) = \displaystyle \frac{\overline{W}(f,k)H(k,t)}{\overline{V}(f,t)},
\end{equation}
$\overline{V}= \overline{W}H$, and $\overline{W}$ is an auxiliary parameter. Since $\sum_k \rho_k(f,t) =1$, we can apply the Jensen's inequality to the convex function $(.)^2$:
\begin{equation}
	[WH](f,t)^2 \leq  \sum_k \rho_k(f,t)  \left( \frac{ W(f,k)H(k,t)}{\rho_k(f,t)} \right)^2,
\end{equation}
which finally leads to the following majorization of the first term of $\mathcal{C}$ in~\eqref{eq:cost_levy}:
\begin{equation}
 \sum_{f,t} \frac{[WH](f,t)^2}{X(f,t)} \leq \sum_{f,t} \frac{\overline{V}(f,t)}{X(f,t)}  \sum_k  \frac{ W(f,k)^2 H(k,t) }{\overline{W}(f,k)}.
\label{eq:cost_C1_levy}
\end{equation}
In a similar fashion, we majorize the second term in~\eqref{eq:cost_levy} ($-\log(.)$ is also a convex function) and therefore we obtain the following auxiliary function $G$:
\begin{multline}
G(W,H,\overline{W}) = \sum_{f,t,k} \frac{ \overline{V}(f,t) H(k,t) }{X(f,t) \overline{W}(f,k)}  W(f,k)^2  \\ - 2 \frac{\overline{W}(f,k)H(k,t)}{\overline{V}(f,t)} \log \left(  W(f,k) \frac{\overline{V}(f,t)}{\overline{W}(f,k)}   \right).
\end{multline}
We then set the partial derivative of $G$ with respect to $W$ to zero, which leads to an update rule on $W$. The update rule on $H$ is obtained in exactly the same way. Thus, the updates are:
\begin{equation}
{\theta} \leftarrow {\theta} \odot a_{\theta}^{\odot 1/2},
\label{eq:mm_update}
\end{equation}
where $a_{W}$ and $a_H$ are given by~\eqref{eq:a_W} and~\eqref{eq:a_H}. A \textsc{MATLAB} implementation of this algorithm can be found at~\cite{Magron}.

\subsection{Estimator of the components}

For a source separation task, it can be useful, once the model parameters are estimated, to derive an estimator $\hat{X}_k$ of the isolated components $X_k$. In a probabilistic framework, a natural estimator is given by the posterior expectation of the source given the mixture $\mathbb{E}(X_k|X)$. It has been shown~\cite{Liutkus2015} that for S$\alpha$S random variables, such an estimator is provided by a generalized Wiener filtering.
%When considering complex isotropic Gaussian variables~\cite{Fevotte2009}, such an estimator is provided by the well-known Wiener filter, which has been shown optimal in a least square sense~\cite{Ephraim1984}. More recently, this result has been extended to S$\alpha$S distributions~\cite{Badeau2014a}, thus providing a theoretical justification to the use of generalized Wiener filtering with fractional spectrograms~\cite{Liutkus2015}.
%In~\cite{Magron2016c}, we prove that this result still holds for the P$\alpha$S family.
One contribution of this paper is to prove that this result still holds for the P$\alpha$S family. This finding extends $\alpha$-Wiener filter for the separation of nonnegative variables. Derivations may be found in the companion technical report for this paper~\cite{Magron2016c}.
For L\'evy-distributed random variables ($\alpha=1/2$), we then have:
\begin{equation}
\hat{X}_k = \mathbb{E}(X_k|X) = \frac{\sigma_k^{\odot 1/2}}{\displaystyle \sum_l \sigma_l^{\odot 1/2}} \odot X = \frac{W_k H_k}{\displaystyle \sum_l W_l H_l} \odot X.
\label{eq:wiener_pas}
\end{equation}
%
%Due to space constraints, we cannot detail here the proof of this result, but a full derivation of this estimator is provided in a supporting document~\cite{Magron2016c}.

\section{Experimental evaluation}
\label{sec:exp}

\subsection{Fitting impulsive noise}

\begin{figure}[t]
\hspace{-0.5cm}
	\includegraphics[scale=0.45]{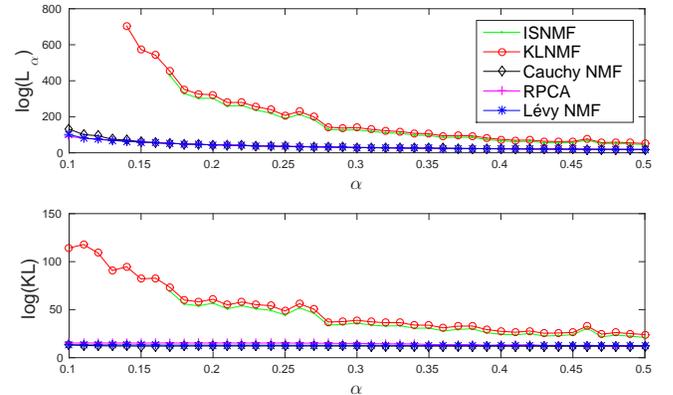}
    \vspace{-0.7cm}
	\caption{Fitting impulsive noise by different algorithms, measured with the $\alpha$-dispersion and KL divergence.}
	\label{fig:levyNMF_toy}
\end{figure}

To test the ability of L\'evy NMF to model impulsive noise, we have generated $5$ components' pairs for $W$ and $H$ by taking the $4$th power of random Gaussian noise, in order to obtain sparse components. The entries of the product $[WH]$, of dimensions $50 \times 50$, were then used as the scale parameters of independent P$\alpha$S random observations, for various values of $\alpha$ in the range $0.1-0.5$: small values of $\alpha$ lead to very impulsive observations. We ran $200$ iterations of the L\'evy, IS~\cite{Fevotte2009}, KL~\cite{Lee1999} and Cauchy~\cite{Liutkus2015a} NMFs, and RPCA~\cite{Candes2011,Huang2012} algorithms with rank $K=5$. To measure the quality of the estimation, we computed the KL divergence and the $\alpha$-dispersion, defined as a function of the data shape parameter $\alpha$:
\begin{equation}
L_{\alpha}=\displaystyle \sum_{f,t} | \sigma(f,t) - \hat{\sigma}(f,t)   |^{1/\alpha},
\end{equation}
where $\sigma^{\odot \alpha} = WH$ contains the synthetic parameters and $\hat{\sigma}^{\odot \alpha} = \hat{W}\hat{H}$ contains the parameters estimated with the different algorithms. Results averaged over $100$ synthetic data runs are presented in Fig.~\ref{fig:levyNMF_toy}. We observe that the L\'evy NMF algorithm shows very similar results to those obtained using RPCA or Cauchy NMF, with slightly better results than these methods for very small values of $\alpha$. The reconstruction quality is considerably better than with ISNMF and KLNMF. Those results demonstrate the potential of the L\'evy NMF model for fitting data which may be very impulsive.

\subsection{Music spectrogram inpainting}

We propose to test the denoising ability of the L\'evy NMF model when the data is corrupted by very impulsive noise. When audio spectrograms are corrupted by such noise, the retrieval of the lost information is known as an~\emph{audio inpainting} task. We consider $6$ guitar songs from the IDMT-SMT-GUITAR~\cite{Kehling2014} database, sampled at $8000$ Hz. The data $X$ is obtained by taking the magnitude spectrogram of the STFT of the mixture signals, computed with a $125$ ms-long Hann window and $75$ $\%$ overlap. The spectrograms are then corrupted with synthetic impulsive noise that represents $10$ $\%$ of the data. We then run $200$ iterations of the algorithms with rank $K=30$ in order to estimate the clean spectrograms.

\begin{figure}[t]
	\centering
	\includegraphics[scale=0.6]{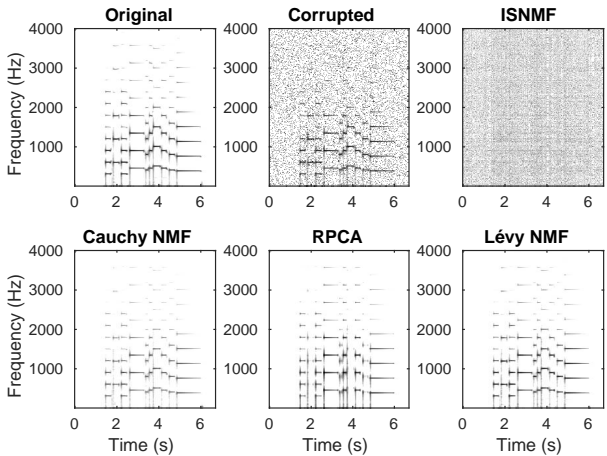}
	\caption{Music spectrogram restoration.}
	\label{fig:levyNMF_guitar_spec}
\end{figure}

We present the obtained spectrograms in Fig.~\ref{fig:levyNMF_guitar_spec} (KLNMF and ISNMF lead to similar results). It appears that the traditional NMF techniques are not able to denoise the data: the estimation of the parameters is deteriorated by the presence of impulsive noise. Conversely, the noise has been entirely removed in the L\'evy NMF estimate. This is confirmed in Table~\ref{tab:levyNMF_guitar_spec} which presents the quality of the estimation measured with the KL divergence between the original and estimated spectrograms averaged over the $6$ audio excerpts. The best results are obtained with L\'evy and Cauchy NMFs.

Remarkably, none of the algorithms used above is informed with any prior knowledge about the location of the noise. As a complementary experiment, we informed ISNMF with this knowledge by incorporating a mask containing the position of the noise into the NMF, resulting into a weighted ISNMF~\cite{Limem2013}. The blind L\'evy NMF still leads to better results than the informed ISNMF. It is thus promising for robust musical applications such as audio inpainting, since it does not require any additional noise modeling or detection technique.

% \begin{table}[t]
% 	\center
% 	\caption{Average Log-KL divergence between clean and estimated music spectrograms.}
% 	\label{tab:levyNMF_guitar_spec}
%     \begin{footnotesize}
% 	\begin{tabular}{|c|c|c|c|c||c|}
% 		\hline
% 		  ISNMF  & KLNMF & Cauchy NMF & RPCA & L\'evy NMF & Weighted ISNMF  \\
% 		\hline
% 		 $9.0$ & $6.2$ & $3.4$ & $3.6$ & $\mathbf{3.2}$  & $3.8$ \\
%          \hline
% 	\end{tabular}
%     \end{footnotesize}
% \end{table}

\begin{table}[t]
	\center
	\caption{Average Log-KL divergence between clean and estimated music spectrograms.}
	\label{tab:levyNMF_guitar_spec}
	\begin{tabular}{|c|c|}
		\hline
         ISNMF & $9.0$ \\
                  \hline
         KLNMF & $6.2$ \\
                  \hline               
         Cauchy NMF & $3.4$ \\
                  \hline
         RPCA & $3.6$ \\
                  \hline
          L\'evy NMF & $\mathbf{3.2}$ \\
		\hline
        \hline
        Weighted ISNMF &  $3.8$ \\
         \hline
	\end{tabular}
\end{table}

\subsection{Application to fluorescence spectroscopy}

%An important problem in chemistry is to identify the concentrations of pure species that compose a mixture.
Fluorescence spectroscopy~\cite{Liu2013} consists in measuring the excitation-emission spectra of a mixture, which is modeled as a the weighted sum of the spectra of the pure species. Performing source separation on these data allows us to identify the pure species composing the mixture and their concentrations. NMF-based methods have shown promising results for this task~\cite{Gobinet2004}, since the data is nonnegative and the NMF model conforms to the underlying assumption of additivity of pure spectra. In this framework, $W$ represents the spectra and $H$ is the concentration matrix.

We propose to apply the L\'evy NMF model to such a task. We compare it with the NMF with Euclidean distance (EuNMF)~\cite{Montcuquet2009} and with KL divergence~\cite{Gobinet2005}. Indeed, other techniques (ISNMF, Cauchy NMF and RPCA) focus on the modeling of complex-valued data, or do not enforce a nonnegative property of the parameters. Thus, it would not be theoretically justified to use those methods in this context.

\begin{figure}[t]
\hspace{-0.5cm}
	\includegraphics[scale=0.45]{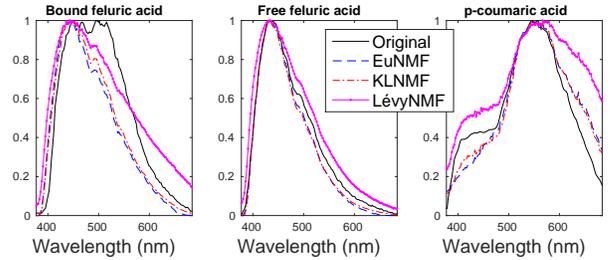}
	\caption{Pure components' spectra learned with several NMF methods.}
	\label{fig:levyNMF_fluor_W}
\end{figure}

Each algorithm uses $50$ iterations of MUR. The dataset is detailed in~\cite{Gobinet2004}. In a nutshell, it consists of $T=400$ emission spectra (with $F=128$ frequency channels) of mixtures of $K=3$ components (bound ferulic acid, free ferulic acid and p-coumaric acid) with unknown concentrations. Both $W$ and $H$ are learned directly from the mixtures. We also have access to the pure spectra of the components. As it is shown in Fig.~\ref{fig:levyNMF_fluor_W}, the L\'evy NMF model seems to be an appropriate tool to learn pure fluorescence spectra from their mixtures. Globally, both Euclidean, KL and L\'evy NMF approximate quite accurately the original pure spectra, though the L\'evy NMF estimate seems to approach the spectra from above.

Finally, we estimate the isolated sources by means of~\eqref{eq:wiener_pas} for the different methods. As a comparison reference, we also learn the concentration matrix $H$ when assuming the pure spectra are known (Oracle case) by means of EuNMF (results obtained with other NMFs are similar). The similarity between the Oracle sources and the estimated sources is measured by means of the correlation, the values of which are presented in Table~\ref{tab:levyNMF_fluor}. The best performance is obtained with the L\'evy NMF method for all sources, which confirms the potential of such a model in various areas of research involving source separation of nonnegative data.

\begin{table}[t]
	\center
	\caption{Correlation (in $\%$) between Oracle and estimated sources.}
	\label{tab:levyNMF_fluor}
	\begin{tabular}{|c||c|c|c|}
		\hline
		    & EuNMF & KLNMF & L\'evy NMF  \\
		\hline
		Bound ferulic acid & $82.8$ & $85.3$ & $\mathbf{87.8}$  \\
		\hline
		Free ferulic acid & $99.5$ & $99.4$ & $\mathbf{99.6}$  \\
		\hline
		p-coumaric acid & $97.3$ & $98.1$ & $\mathbf{98.4}$  \\
		\hline
	\end{tabular}
\end{table}

\section{Conclusion}
\label{sec:conclu}

In this paper, we introduced the L\'evy NMF model, which structures the dispersion parameters of P$\alpha$S distributed sources when $\alpha=1/2$.
%We proposed to estimate the model parameters in an ML sense, and we also derived an estimator of the isolated sources, which extends the validity of the generalized Wiener filtering to the family of P$\alpha$S distributions.
Experiments have shown the potential of this model for robustly decomposing realistic nonnegative data. 

Such a model could be useful in many other fields where the source separation issue frequently occurs and where the L\'evy distribution finds applications, such as optics~\cite{Rogers2008} or geomagnetic field analysis~\cite{Carbone2006}. Future work could focus on novel estimation techniques for the L\'evy NMF model, using for instance a MAP estimator, which would permit us to incorporate some prior knowledge about the parameters~\cite{Bertin2010}. Besides, drawing on~\cite{Simsekli2015}, the family of techniques based on MCMC could be useful to estimate the parameters of any P$\alpha$S distribution.
%even when the PDF cannot be written in closed form.
%Alternatively, one could extend the L\'evy NMF model to the family of inverted gamma (IG) distributions, of which L\'evy is one particular case. IG distributions are tractable, but they are not additive in general, which would require to design some variational inference schemes~\cite{Kounades-Bastian2016} for estimating such models.
Alternatively, one could extend the L\'evy NMF model to the family of inverted gamma (IG) of which it is a special case. Although it would be losing additivity and thus the theoretical foundation for $\alpha$-Wiener filtering, this would allow for convenient analytical derivations thanks to tractable likelihood functions. This strategy is reminiscent of recent work \cite{Yoshii2016} where the tractable student-t distribution is used instead of the symmetric $\alpha$-stable one.

\section*{Acknowledgment}

The authors would like to thank Cyril Gobinet from University of Reims Champagne-Ardenne, France, for providing them the fluorescence spectroscopy data used in this study.

\ifCLASSOPTIONcaptionsoff
  \newpage
\fi




\newpage
\bibliographystyle{IEEEtran}
\bibliography{IEEEabrv,references}

\begin{thebibliography}{10}
\providecommand{\url}[1]{#1}
\csname url@samestyle\endcsname
\providecommand{\newblock}{\relax}
\providecommand{\bibinfo}[2]{#2}
\providecommand{\BIBentrySTDinterwordspacing}{\spaceskip=0pt\relax}
\providecommand{\BIBentryALTinterwordstretchfactor}{4}
\providecommand{\BIBentryALTinterwordspacing}{\spaceskip=\fontdimen2\font plus
\BIBentryALTinterwordstretchfactor\fontdimen3\font minus
  \fontdimen4\font\relax}
\providecommand{\BIBforeignlanguage}[2]{{%
\expandafter\ifx\csname l@#1\endcsname\relax
\typeout{** WARNING: IEEEtran.bst: No hyphenation pattern has been}%
\typeout{** loaded for the language `#1'. Using the pattern for}%
\typeout{** the default language instead.}%
\else
\language=\csname l@#1\endcsname
\fi
#2}}
\providecommand{\BIBdecl}{\relax}
\BIBdecl

\bibitem{Pauca2004}
V.~P. Pauca, F.~Shahnaz, M.~W. Berry, and R.~J. Plemmons, ``{Text mining using
  non-negative matrix factorizations},'' in \emph{{Proc. SIAM international
  conference on data mining}}, Lake Buena Vista, Florida, USA, January 2004,
  pp. 452--456.

\bibitem{Guillamet2002}
D.~Guillamet and J.~Vitria, ``Classifying faces with nonnegative matrix
  factorization,'' in \emph{Proc. the Catalan conference for artificial
  intelligence}, Castell\'{o}n, Spain, October 2002, pp. 24--31.

\bibitem{Sajda2004}
P.~Sajda, S.~Du, T.~R. Brown, R.~Stoyanova, D.~C. Shungu, X.~Mao, and L.~C.
  Parra, ``Nonnegative matrix factorization for rapid recovery of constituent
  spectra in magnetic resonance chemical shift imaging of the brain,''
  \emph{IEEE Transactions on Medical Imaging}, vol.~23, no.~12, pp. 1453--1465,
  December 2004.

\bibitem{Lee1999}
D.~D. Lee and H.~S. Seung, ``{Learning the parts of objects by non-negative
  matrix factorization},'' \emph{Nature}, vol. 401, no. 6755, pp. 788--791,
  1999.

\bibitem{Smaragdis2003}
P.~Smaragdis and J.~C. Brown, ``{Non-negative matrix factorization for
  polyphonic music transcription},'' in \emph{{Proc. IEEE Workshop on
  Applications of Signal Processing to Audio and Acoustics (WASPAA)}}, New
  Paltz, NY, USA, October 2003.

\bibitem{Liu2013}
P.~Liu, X.~Zhou, Y.~Li, M.~Li, D.~Yu, and J.~Liu, ``{The application of
  principal component analysis and non-negative matrix factorization to analyze
  time-resolved optical waveguide absorption spectroscopy data},''
  \emph{Analytical Methods}, vol.~5, pp. 4454--4459, 2013.

\bibitem{Fevotte2009}
C.~F{\'e}votte, N.~Bertin, and J.-L. Durrieu, ``{Nonnegative matrix
  factorization with the {Itakura-Saito} divergence: With application to music
  analysis},'' \emph{Neural computation}, vol.~21, no.~3, pp. 793--830, March
  2009.

\bibitem{Schmidt2008}
M.~N. Schmidt and H.~Laurberg, ``{Nonnegative Matrix Factorization with
  Gaussian Process Priors},'' \emph{Computational Intelligence and
  Neuroscience}, vol. 2008, no.~3, pp. 3:1--3:10, January 2008.

\bibitem{Virtanen2008}
T.~Virtanen, A.~T. Cemgil, and S.~Godsill, ``{Bayesian extensions to
  non-negative matrix factorisation for audio signal modelling},'' in
  \emph{{Proc. IEEE International Conference on Acoustics, Speech and Signal
  Processing (ICASSP)}}.\hskip 1em plus 0.5em minus 0.4em\relax Las Vegas, NV,
  USA: IEEE, May 2008, pp. 1825--1828.

\bibitem{Cemgil2009}
A.~T. Cemgil, ``Bayesian inference for nonnegative matrix factorisation
  models,'' \emph{Computational Intelligence and Neuroscience}, vol. 2009,
  2009, article ID 785152, 17 pages.

\bibitem{Bertin2010}
N.~Bertin, R.~Badeau, and E.~Vincent, ``Enforcing harmonicity and smoothness in
  {Bayesian} non-negative matrix factorization applied to polyphonic music
  transcription,'' \emph{IEEE Transactions on Audio, Speech and Language
  Processing}, vol.~18, no.~3, pp. 538--549, March 2010.

\bibitem{Virtanen2007}
T.~Virtanen, ``Monaural sound source separation by nonnegative matrix
  factorization with temporal continuity and sparseness criteria,'' \emph{IEEE
  Transactions on Audio, Speech, and Language Processing}, vol.~15, no.~3, pp.
  1066--1074, March 2007.

\bibitem{Samoradnitsky1994}
G.~Samoradnitsky and M.~S. Taqqu, \emph{{Stable non-Gaussian random processes:
  stochastic models with infinite variance}}.\hskip 1em plus 0.5em minus
  0.4em\relax CRC Press, 1994.

\bibitem{Godsill1999}
S.~Godsill and E.~E. Kuruoglu, ``Bayesian inference for time series with
  heavy-tailed symmetric $\alpha$-stable noise processes,'' \emph{Applications
  of Heavy Tailed Distributions in Economics, Engineering and Statistics (Heavy
  Tails 99)}, pp. 3--5, 1999.

\bibitem{Bassiou2013}
N.~Bassiou, C.~Kotropoulos, and E.~Koliopoulou, ``Symmetric $\alpha$-stable
  sparse linear regression for musical audio denoising,'' in \emph{Proc.
  International Symposium on Image and Signal Processing and Analysis
  (ISPA)}.\hskip 1em plus 0.5em minus 0.4em\relax Trieste, Italy: IEEE,
  September 2013, pp. 382--387.

\bibitem{Liutkus2015}
A.~Liutkus and R.~Badeau, ``{Generalized Wiener filtering with fractional power
  spectrograms},'' in \emph{{Proc. the IEEE International Conference on
  Acoustics, Speech and Signal Processing (ICASSP)}}, Brisbane, Australia,
  April 2015.

\bibitem{Liutkus2015a}
A.~Liutkus, D.~Fitzgerald, and R.~Badeau, ``{Cauchy Nonnegative Matrix
  Factorization},'' in \emph{{Proc. IEEE Workshop on Applications of Signal
  Processing to Audio and Acoustics (WASPAA)}}, New Paltz, NY, USA, October
  2015.

\bibitem{Simsekli2015}
U.~Simsekli, A.~Liutkus, and A.~T. Cemgil, ``Alpha-stable matrix
  factorization,'' \emph{IEEE Signal Processing Letters}, vol.~22, no.~12, pp.
  2289--2293, 2015.

\bibitem{Nolan2015}
J.~P. Nolan, \emph{Stable Distributions - Models for Heavy Tailed Data}.\hskip
  1em plus 0.5em minus 0.4em\relax Boston: Birkhauser, 2015, in progress,
  Chapter 1 online at academic2.american.edu/$\sim$jpnolan.

\bibitem{Hunter2004}
D.~R. Hunter and K.~Lange, ``{A tutorial on {MM} algorithms},'' \emph{The
  American Statistician}, vol.~58, no.~1, pp. 30--37, 2004.

\bibitem{Fevotte2011}
C.~F{\'e}votte and J.~Idier, ``{Algorithms for nonnegative matrix factorization
  with the beta-divergence},'' \emph{Neural Computation}, vol.~23, no.~9, pp.
  2421--2456, September 2011.

\bibitem{Fevotte2011a}
C.~F{\'e}votte, ``{Majorization-minimization algorithm for smooth Itakura-Saito
  nonnegative matrix factorization},'' in \emph{Proc. IEEE International
  Conference on Acoustics, Speech and Signal Processing (ICASSP)}.\hskip 1em
  plus 0.5em minus 0.4em\relax Prague, Czech Republic: IEEE, May 2011, pp.
  1980--1983.

\bibitem{Magron}
``{Webpage},''
  "\url{http://www.perso.telecom-paristech.fr/magron/ressources/levyNMF.m}".

\bibitem{Magron2016c}
P.~Magron, R.~Badeau, and A.~Liutkus, ``{Generalized Wiener filtering for
  positive alpha-stable random variables},'' T{\'e}l{\'e}com ParisTech, Tech.
  Rep. D004, 2016.

\bibitem{Candes2011}
E.~J. Cand{\`e}s, X.~Li, Y.~Ma, and J.~Wright, ``{Robust principal component
  analysis?}'' \emph{Journal of the ACM (JACM)}, vol.~58, no.~3, p.~11, 2011.

\bibitem{Huang2012}
P.-S. Huang, S.~D. Chen, P.~Smaragdis, and M.~Hasegawa-Johnson,
  ``{Singing-voice separation from monaural recordings using robust principal
  component analysis},'' in \emph{{Proc. IEEE International Conference on
  Acoustics, Speech and Signal Processing (ICASSP)}}, Kyoto, Japan, March 2012,
  pp. 57--60.

\bibitem{Kehling2014}
C.~Kehling, A.~Jakob, D.~Christian, and S.~Gerald, ``{Automatic tablature
  transcription of electric guitar recordings by estimation of score- and
  instrument-related parameters},'' in \emph{{Proc. the International
  Conference on Digital Audio Effects (DAFx)}}, Erlangen, Germany, September
  2014, p.~8.

\bibitem{Limem2013}
A.~Limem, G.~Delmaire, M.~Puigt, G.~Roussel, and D.~Courcot, ``Non-negative
  matrix factorization using weighted beta divergence and equality constraints
  for industrial source apportionment,'' in \emph{Proc. the IEEE International
  Workshop on Machine Learning for Signal Processing (MLSP)}.\hskip 1em plus
  0.5em minus 0.4em\relax Southampton, United Kingdom: IEEE, September 2013,
  pp. 1--6.

\bibitem{Gobinet2004}
C.~Gobinet, E.~Perrin, and R.~Huez, ``Application of non-negative matrix
  factorization to fluorescence spectroscopy,'' in \emph{{Proc. European Signal
  Processing Conference (EUSIPCO)}}, Vienna, Austria, September 2004.

\bibitem{Montcuquet2009}
A.-S. Montcuquet, L.~Herve, L.~Guyon, J.-M. Dinten, and J.~I. Mars,
  ``Non-negative matrix factorization: A blind sources separation method to
  unmix fluorescence spectra,'' in \emph{Proc. the Workshop on Hyperspectral
  Image and Signal Processing: Evolution in Remote Sensing (WHISPERS)}.\hskip
  1em plus 0.5em minus 0.4em\relax Grenoble, France: IEEE, August 2009, pp.
  1--4.

\bibitem{Gobinet2005}
C.~Gobinet, A.~Elhafid, V.~Vrabie, R.~Huez, and D.~Nuzillard, ``About
  importance of positivity constraint for source separation in fluorescence
  spectroscopy,'' in \emph{{Proc. European Signal Processing Conference
  (EUSIPCO)}}.\hskip 1em plus 0.5em minus 0.4em\relax Antalya, Turkey: IEEE,
  September 2005, pp. 1--4.

\bibitem{Rogers2008}
G.~L. Rogers, ``Multiple path analysis of reflectance from turbid media,''
  \emph{Journal of the Optical Society of America}, vol.~25, no.~11, pp.
  2879--2883, 2008.

\bibitem{Carbone2006}
V.~Carbone, L.~Sorriso-Valvo, A.~Vecchio, F.~Lepreti, P.~Veltri, P.~Harabaglia,
  and I.~Guerra, ``Clustering of polarity reversals of the geomagnetic field,''
  \emph{Physical review letters}, vol.~96, no.~12, March 2006, article no.
  128501.

\bibitem{Yoshii2016}
K.~Yoshii, K.~Itoyama, and M.~Goto, ``Student's t nonnegative matrix
  factorization and positive semidefinite tensor factorization for
  single-channel audio source separation,'' in \emph{Proc. of the IEEE
  International Conference on Acoustics, Speech, and Signal Processing
  (ICASSP)}.\hskip 1em plus 0.5em minus 0.4em\relax IEEE, 2016.

\end{thebibliography}

\end{document}